# Mode conversion in nonlinear waveguides stimulated by the longitudinal bi-harmonic refractive index modulation

Yaroslav V. Kartashov[1] and Victor A. Vysloukh[2]

[1]Institute of Spectroscopy, Russian Academy of Sciences, Troitsk, Moscow Region, 142190, Russia

[2]Departamento de Fisica y Matematicas, Universidad de las Americas – Puebla, Santa Catarina Martir, 72820, Puebla, Mexico

We study specific features of resonant mode conversion in nonlinear waveguides stimulated by the bi-harmonic longitudinal modulation of its parameters, which includes changes of the waveguide depth as well as its bending (in the one-dimensional case) or spiraling (in the two-dimensional case). We demonstrate the possibility of simultaneous excitation of higher-order modes of different parities and topologies with controllable energy weights. The output mode composition is highly sensitive to the variation in the input power and detuning from the resonant modulation frequency.

Quantum revivals stimulated by periodic external fields, first introduced by I. Rabi in his pioneering work [1], were extensively studied in atomic, molecular, and other physical settings (see [2] for a recent review and references therein). Upon revivals the states occupying initially only one energy level of the system spread over several energy levels, but later the spreading is reversed, and the system comes back to its initial state. Such revivals occur periodically. Optical systems (such as multimode waveguides, waveguide arrays, photonic lattices or photonic crystals) provide a unique laboratory for the investigation of analogs of quantum revivals, since in such systems the direct monitoring of spatial intensity distributions and sometimes of propagation dynamics inside the sample is possible. Various resonant effects were already demonstrated experimentally in such settings [3-7]. An extensive recent survey of this topic can be found in Ref. [8].

In practice, optical revivals (or stimulated mode conversion) have been realized in optical fibers, subjected to periodic mechanical stress, and in long-period photoinduced gratings [9-12]. Conversion of guided modes, stimulated by shallow longitudinal refractive index modulations, was predicted in multimode waveguides [13] and demonstrated experimentally in photonic lattices [14]. Challenging problem, which is of special interest for numerous applications, is the realization of stimulated transition between states with trivial phase distributions and vortex-carrying states that have essentially different topologies. Different approaches to the transfer of orbital angular momentum from a spiraling or stressed fiber-optic waveguide to a light beam were elaborated and presented [17-19]. In particular, the generation, cancellation, and revivals of different vortex states, which are mediated by shallow refractive index modulations, were reported recently [20]. Similar effects may be observed in Bose-Einstein condensates held in optical traps and driven by resonant fields [21,22]. The control of the evolution and modal composition of the field is important in various areas of physics, especially when the field contains topological dislocations (see for example [23,24] and references therein).

In this Letter we exploit new prospects for stimulated mode conversion opened by application of bi-harmonic modulation of parameters of multimode nonlinear waveguides. To the best of our knowledge the revivals due to bi-harmonic modulation were never addressed previously. We start our analysis with one-dimensional linear Gaussian waveguide and show how periodic longitudinal modulation of linear refractive index, combined with harmonic bending, induces simultaneous conversion of fundamental mode into dipole and tripole modes, as well as different cascading transformations. The mode composition of the output light pattern (or energy weights of guided modes) can be efficiently controlled by tuning frequencies of bending and refractive index modulation. In the nonlinear regime, the output mode composition is sensitive to small variations of the input power. We also uncovered that two-dimensional settings enrich opportunities for linear and nonlinear mode conversion, since in such settings one can achieve simultaneous conversion into vortex and ring-like modes.

We describe the propagation of light beams along the $\xi$-axis of transparent dielectric with cubic nonlinearity and nontrivial bi-harmonic spatial modulation of the refractive index by the nonlinear Schrödinger equation for the normalized light field amplitude $q(\eta,\zeta,\xi)$:

$$i\frac{\partial q}{\partial \xi} = -\frac{1}{2}\Delta q - pR(\eta,\zeta,\xi)q + \sigma |q|^2 q. \qquad (1)$$

In this equation the transverse coordinates $\eta,\zeta$ are normalized to the characteristic scale $r_0$; the Laplace operator has the form $\Delta = \partial^2/\partial\eta^2 + \partial^2/\partial\zeta^2$ in the two-dimensional (2D) case and reduces to $\Delta = \partial^2/\partial\eta^2$ in the one-dimensional (1D) setting. The propagation distance $\xi$ is normalized by the diffraction length $kr_0^2$, where $k = 2\pi n/\lambda$ is the wavenumber; the nonlinear parameter $\sigma > 0$ ($\sigma < 0$) corresponds to defocusing (focusing) nonlinearity, and the guiding parameter $p = k^2 r_0^2 \delta n/n$ is proportional to a small variation $\delta n$ of the refractive index $n$. The function $R(\eta,\zeta,\xi)$ describes the refractive index profile of the waveguide. In the case of fused silica (at $\lambda = 0.633$ $\mu$m and $r_0 = 5\mu$m) the parameter $p = 1$ corresponds to the refractive index contrast $\delta n \simeq 0.7\times10^{-4}$, while propagation distance $\xi = 1$ is equivalent to $\approx 0.35$ mm of real propagation distance. Notice that Eq.(1) disregards backward scattering, which vanishes when the longitudinal refractive index modulation is slow in comparison with the wavelength $\lambda$ and when the refractive index contrast is low, i.e. $\delta n/n \ll 1$.

Further, we consider two different combinations of the refractive index modulation. In the 1D case the modulation of waveguide depth will be mixed with its harmonic bending:

$$R(\eta,\xi) = [1 + \mu_d \sin(\Omega_d \xi)] \exp\{-[\eta - \mu_b \sin(\Omega_b \xi)]^2/a^2\}, \qquad (2)$$

while in the 2D case, the modulation of depth will be accompanied by periodic spiraling of the waveguide in the direction of propagation:

$$R(\eta,\zeta,\xi) = [1 + \mu_d \sin(\Omega_d \xi)] \exp\{-[\eta + \mu_r \sin(\Omega_r \xi)]^2/a^2 - [\zeta - \mu_r \cos(\Omega_r \xi)]^2/a^2\}. \qquad (3)$$

In Eqs. (2) and (3) $a$ is the waveguide width, $\mu_d, \Omega_d$ stand for the amplitude and spatial frequency of the depth modulation, $\mu_b, \Omega_b$ are the amplitude and frequency of waveguide bending in the one-dimensional case, and $\mu_r, \Omega_r$ are the amplitude and frequency of spiraling in the two-dimensional case.

In what follows, we focus our attention on the case of waveguides that support at least three guided modes, and study numerically how combined and nearly resonant modulation of the waveguide depth and bending (in the 1D case) or spiraling (in the 2D case) stimulates mode coupling and conversion, and how Kerr-type cubic nonlinearity affects this process. Notice that in a linear limit ($\sigma = 0$) the evolution of light in such a system will demonstrate features similar to evolution of populations in a three-level quantum system under the influence of bi-harmonic pumping.

For preliminary insight into dynamics of this system in the linear limit and in the 1D case, it is instructive to use the standard technique of analysis of resonant mode coupling, based on the modal expansion: $q(\eta,\xi) = c_0(\xi)w_0(\eta)e^{ib_0\xi} + c_1(\xi)w_1(\eta)e^{ib_1\xi} + c_2(\xi)w_2(\eta)e^{ib_2\xi}$, which, after the substitution of the light field in such form into Eq. (1), yields the system of equations describing the evolution of the complex mode amplitudes $c_j(\xi)$ that is reminiscent of the system describing the evolution of populations of different levels under the effect of external field in quantum system:

$$\frac{dc_0}{d\xi} = i\mu_d S_{00} \sin(\Omega_d \xi) c_0 - \frac{1}{2}\mu_b A_{01} e^{i(\Omega_b - \Omega_{01})\xi} c_1 + \frac{1}{2}\mu_d S_{02} e^{i(\Omega_d - \Omega_{02})\xi} c_2,$$
$$\frac{dc_1}{d\xi} = \frac{1}{2}\mu_b A_{01} e^{-i(\Omega_b - \Omega_{01})\xi} c_0 + i\mu_d S_{11} \sin(\Omega_d \xi) c_1 - \frac{1}{2}\mu_b A_{12} e^{i(\Omega_b - \Omega_{12})\xi} c_2, \quad (4)$$
$$\frac{dc_2}{d\xi} = -\frac{1}{2}\mu_d S_{02} e^{-i(\Omega_d - \Omega_{02})\xi} c_0 + \frac{1}{2}\mu_b A_{12} e^{-i(\Omega_d - \Omega_{12})\xi} c_1 + i\mu_d S_{22} \sin(\Omega_d \xi) c_2,$$

where the exchange integrals are introduced as

$$S_{jk} = \int_{-\infty}^{\infty} w_j(\eta) p R_0(\eta) w_k(\eta) d\eta,$$
$$A_{jk} = \int_{-\infty}^{\infty} w_j(\eta) p R_0'(\eta) w_k(\eta) d\eta, \quad (5)$$

$w_k(\eta)$ are the shapes of guided modes with different number of nodes $k$, $b_k$ are the propagation constants of different modes, $\Omega_{jk} = b_j - b_k$ are the transitions frequencies, $R_0 = R|_{\xi=0}$ and $R_0' = \partial R/\partial \eta|_{\xi=0}$ are the profile and derivative of the profile of static waveguide. One can see from Eq. (4) that longitudinal modulation leads to resonant coupling of different modes. Notice that due to depth modulation, the exchange of energy is possible only between modes with equal parity $w_0 \leftrightarrow w_2$, since otherwise the exchange integral $S_{jk}$ vanishes for symmetric refractive index distribution $R_0(\eta) = R_0(-\eta)$. On the other hand, bending introduces interaction between modes of different parity ($w_0 \leftrightarrow w_1$ and $w_1 \leftrightarrow w_2$) because corresponding integrals $A_{jk}$ contain anti-symmetric function $R_0'(\eta) = -R_0'(-\eta)$ and are nonzero in this case. When depth modulation and bending act simultaneously, the power from $w_0$ mode will flow simultaneously into modes $w_1$ and $w_2$. It should be stressed that under resonant conditions ($\Omega_b \simeq \Omega_{01}$ and $\Omega_d \simeq \Omega_{02}$) corresponding to most effective $w_0 \leftrightarrow w_1$, $w_0 \leftrightarrow w_2$ coupling the transitions between exited states $w_1 \leftrightarrow w_2$ are weak due to the presence of rapidly oscillating exponential factors $\exp[i(\Omega_b - \Omega_{12})]$ and $\exp[-i(\Omega_d - \Omega_{12})]$ in corresponding terms in Eqs. (4). Diagonal terms in the right part of Eqs.(4) stand for the trivial periodic phase modulation of the modal weights $c_j(\xi)$ due to the depth modulation. Non-diagonal terms determine the rates of inter-modal energy exchange. For instance, at resonant conditions and at $\mu_d \to 0$ the amplitudes $c_0$, $c_1$ oscillate with the frequency $(1/2)\mu_b A_{01}$, while at

$\mu_b \to 0$ Eqs.(4) predicts oscillations of $c_0, c_2$ amplitudes with the frequency $(1/2)\mu_d S_{02}$. Obviously, bi-harmonic modulation opens broad prospects for the control of modal structure of the output field and realization of various cascade transformations.

While the system of Eqs. (4) for mode amplitudes, obtained in the frames of coupled-mode approach, allows to get main qualitative predictions about system dynamics, we further resort to numerical integration of the full model (1)-(3). This is mainly due to the fact that idealized system (4) does not account for radiative losses that unavoidably occur upon longitudinal modulation of real refractive index landscape and that may become especially pronounced for higher-order modes. Radiative losses lead to continuous leakage of power from the system and impose the restrictions on the maximal possible modulation depth, which are absent in (4). Moreover, Eqs. (4) account only for direct resonances and do not take into account the possibility of parametric processes (one example with parametric generation of vortex modes for forbidden direct resonance is provided in Ref. [20]). Finally, Eqs. (1)-(3) allow most accurate inclusion of material nonlinearity that affects mode shapes and their propagation constants.

The process of mode conversion in the one-dimensional case is illustrated in Figs. (1)-(3). The results were obtained by direct solution of evolution Eq. (1). The static $(\mu_d, \mu_b = 0)$ one-dimensional Gaussian waveguide supports at $a=2$, $p=2.3$ only three guided modes: fundamental $w_0$, dipole $w_1$, and tripole $w_2$ ones, with propagation constants $b_0 = 1.8110$, $b_1 = 0.9393$, and $b_2 = 0.3042$, correspondingly. In the limit of weak nonlinearity and for shallow modulation ($\mu_d, \mu_b \ll 1$, $\sigma \ll 1$), any solution of Eq.(1) can be adequately represented by the linear superposition of guided modes, and the contribution of each particular mode to the conserved total energy flow $U = \int_{-\infty}^{\infty} |q|^2 d\eta$ can be characterized by its energy weight, which is introduced as $\nu_k(\xi) = |c_k(\xi)|^2$, where $k=0,1,2$. Notice that the modes are normalized in such a way that $\int_{-\infty}^{\infty} w_k^2 d\eta = 1$.

Figure 1(a) illustrates typical evolution of the energy weights $\nu_k(\xi) = |c_k(\xi)|^2$, which were extracted from the instantaneous field distributions $c_k(\xi) = \int_{-\infty}^{\infty} q(\eta, \xi) w_k(\eta) d\eta$, for the case when only fundamental mode $w_0$ is launched into waveguide at $\xi = 0$ under the condition of exact resonances: $\Omega_b = \Omega_{01}$ and $\Omega_d = \Omega_{02}$. Corresponding evolution dynamics is illustrated in Fig. 2(a). Here we recall that the bending is responsible for conversion of the fundamental mode into the dipole one $w_0 \to w_1$, while the depth modulation stimulates transition between fundamental and tripole modes $w_0 \to w_2$. The modulation amplitudes $\mu_b, \mu_d$ in Fig. 1(a) were adjusted to provide equal energy weights $\nu_1 \approx \nu_2 \approx 0.5$ of dipole and tripole modes at the same distance $\xi \approx 232$. Notice that small-amplitude oscillations on $\nu_1(\xi), \nu_2(\xi)$ dependencies are linked with a weak energy exchange between excited states $w_1 \leftrightarrow w_2$. Notice that energy exchange is periodic, i.e. nearly all power returns into fundamental mode after one complete energy exchange period. Bi-harmonic modulation allows to realize cascade processes $w_1 \to w_0 \to w_2$ and $w_2 \to w_0 \to w_1$, when transition between initial and final states occurs via "intermediate" state, as shown in Figs. 1(b) and 1(c). Figures 2(b) and 2(c) illustrate evolution dynamics in the $(\eta, \xi)$ plane encountered upon such cascading transformations. It should be pointed out that inclusion of the longitudinal modulation on additional, third frequency may further enrich evolution dynamics of the system because it may directly couple for example $w_1$ and $w_2$ modes and this coupling may accompany cascading processes shown above. At the same time, already bi-harmonic modulation is sufficient to achieve any desired combination of all three modes at the output.

For further discussion, we introduce relative frequency detuning of bending and depth modulation frequencies $\delta\Omega_b = (\Omega_b - \Omega_{01})/\Omega_{01}$ and $\delta\Omega_d = (\Omega_d - \Omega_{02})/\Omega_{02}$ from their resonant values and focus our attention on the resonant features of the bi-harmonic modulation. Figures 3(a) and 3(b) illustrate resonant curves for simultaneous conversion of the fundamental mode into the dipole and tripole ones $(w_0 \to w_1, w_2)$ under the condition of exact resonance

$\delta\Omega_d = 0$ between fundamental and tripole modes and for various detuning $\delta\Omega_b$ from resonance between fundamental and dipole modes (the modulation amplitudes $\mu_d = 0.0333$, $\mu_b = 0.0167$ were fixed). One can see that already a few percent detuning of the bending frequency from the resonant value remarkably reduces the maximal possible weight $\nu_1^{\max}$ (determined over all distances $\xi$) of the dipole mode, thereby increasing maximal possible weight of the tripole mode $\nu_2^{\max}$. Interestingly, the distances $\xi_1^{\max}, \xi_2^{\max}$ at which weights of the dipole and tripole modes reach their maximal values may strongly differ for $\delta\Omega_b \neq 0$ and they become nearly equal only in exact resonance. Figure 2(c) exemplifies the dependences of the maximal energy weights on the amplitude of bending $\mu_b$ at fixed $\mu_d = 0.0333$. One can see that by varying one of modulation depths one can control the modal content of the field. Increasing $\mu_b$ results in growth and subsequent saturation of $\nu_1^{\max}$. Notice that variation of $\delta\Omega_d$ at fixed $\delta\Omega_b = 0$ leads to a similar result. The main conclusion is that mode composition of the output pattern is very sensitive to small detuning of the bending or depth modulation frequency.

We found that cubic nonlinearity (focusing or defocusing) drastically affects the process of mode conversion under the action of bi-harmonic modulation (Fig. 4). Nonlinearity introduces additional shift of propagation constants of eigenmodes of the waveguide, which is different for different modes. As a result, the modulation and bending frequencies corresponding to the most effective resonant conversion shift with increase of nonlinearity coefficient $\sigma$ [see Figs. 4(a) and 4(b) showing resonance curves for different $\sigma$ values]. The value of this shift is approximately proportional to the modulus of nonlinear coefficient $\sigma$: resonance frequency increases for focusing nonlinearity and decreases for defocusing nonlinearity. Naturally, this shift radically changes modal content of the output field at fixed propagation distance [see Fig.4(c) illustrating weights of the dipole and tripole modes at $\xi = 232$ as functions of nonlinearity coefficient]. For example, the presence of small focusing nonlinearity ($\sigma = -0.15$) completely eliminates tripole mode $w_2$ from the output field distribution, while in the linear medium at the same distance this mode has weight $\nu_1 \approx \nu_2 \approx 0.5$ nearly equal to the weight of the dipole mode. The width of resonance in terms of nonlinearity coefficient is much smaller for higher-order modes than for lower-order ones. Figures 2(d)-2(f) illustrate dynamics of nonlinear mode conversion at different levels of focusing nonlinearity. This high sensitivity of the output mode composition to the input power variation may be also useful for practical applications.

Even richer opportunities for linear and nonlinear mode conversion under the action of bi-harmonic modulation appear in two-dimensional settings, where one can achieve simultaneous transition between modes with trivial phase distributions and vortex-carrying modes. Figure 5(a) shows eigenmodes $w_k^m$ (here $k$ is the number of radial nodes and $m$ is the topological charge) of static 2D Gaussian waveguide with $a=1$, $p=10$. The single-hump fundamental mode $w_0^0$ with propagation constant $b_0^0 = 6.042$ is followed by the vortex mode $w_0^1$ with $b_0^1 = 2.685$, and then by the ring-like mode $w_1^0$ with trivial phase and propagation constant $b_1^0 = 0.636$. Simultaneous modulation of the waveguide depth and its spiraling at resonant $\Omega_d = 5.406$ and $\Omega_r = 3.357$ lead to the excitation of both higher-order modes by the fundamental mode [see Fig. 6(d) with output intensity distribution in this case, while separate excitation of vortex and ring-like modes by single-frequency modulation is shown in Figs. 6(b) and 6(c)]. If modulation and spiraling amplitudes are chosen properly, then energy weighs $\nu_k^m$ defined as $\nu_k^m(\xi) = |c_k^m(\xi)|^2$, where $c_k^m(\xi) = \iint q^*(r,\phi,\xi) w_k^m(r) \exp(im\phi) dr d\phi$ is the modal amplitude, $r, \phi$ are the radial and angular variables in the $(\eta, \zeta)$ plane, of two higher-order modes may become equal ($\nu_0^1 \approx \nu_1^0 \approx 0.5$) at the same distance (for $\mu_r = 0.002$ and $\mu_d = 0.0081$ it is given by $\xi = 181$). Just like in the 1D case, the variation in one of the modulation frequencies results in notable modifications of the maximal possible weights of both modes. Addition of focusing or defocusing nonlinearity allows to change mode content too, because the width of resonance in $\sigma$ is much narrower for the ring-like mode than for vortex-carrying one [see Fig. 5(b)]. In particular, gradual increase of $\sigma$ may lead to complete elimination of ring-like mode from the output distribution [Figs. 6(d)-6(f)].

Summarizing, we studied resonant mode conversion in linear and nonlinear waveguides under the influence of bi-harmonic longitudinal modulation of its parameters. Such modulation allows simultaneous excitation with controllable weights of modes having different parity. High sensitivity of the mode content of the output field to modulation frequencies and nonlinearity allows drastic modifications of the output intensity distributions.

# References

1. I. Rabi, "On the process of space quantization," Phys. Rev. 49, 324 (1936).
2. R. W. Robinett, "Quantum wave packet revivals," Physics Reports 392, 1 (2004).
3. Y. B. Ovchinnikov, "Revivals of light in a planar metal waveguide," Opt. Commun. 182, 35 (2000).
4. S. Longhi, M. Marangoni, M. Lobino, R. Ramponi, P. Laporta, E. Cianci, and V. Foglietti, "Observation of dynamical localization in periodically curved waveguide arrays," Phys. Rev. Lett. 96, 243901 (2006).
5. R. Iyer, J. S. Aitchison, J. Wan, M. M. Dignam, and C. M. de Sterke, "Exact dynamic localization in curved AlGaAs optical waveguide arrays," Opt. Express **15**, 3212 (2007); F. Dreisow, M. Heinrich, A. Szameit, S. Doring, S. Nolte, A. Tunnermann, S. Fahr, and F. Lederer, "Spectral resolved dynamic localization in curved fs laser written waveguide arrays," Opt. Express **16**, 3474 (2008).
6. S. Longhi, "Optical realization of multilevel adiabatic population transfer in curved waveguide arrays," Phys. Lett. A 359, 166 (2006).
7. K. Staliunas and R. Herrero, "Nondiffractive propagation of light in photonic crystals," Phys. Rev. E **73**, 016601 (2006); S. Longhi and K. Staliunas, "Self-collimantion and self-imaging effects in modulated waveguide arrays," Opt. Commun. **281**, 4343 (2008); A. Szameit, Y. V. Kartashov, F. Dreisow, M. Heinrich, T. Pertsch, S. Nolte, A. Tunnermann, V. A. Vysloukh, F. Lederer, and L. Torner, "Inhibition of light tunneling in waveguide arrays," Phys. Rev. Lett. **102**, 153901 (2009).
8. I. L. Garanovich, S. Longhi, A. A. Sukhorukov, and Y. S. Kivshar, "Light propagation and localization in modulated photonic lattices and waveguides," Phys. Rep. 518, 1 (2012).
9. K. O. Hill, B. Malo, K. A. Vineberg, F. Bilodeau, D. C. Johnson, and I. Skinner, "Efficient mode conversion in telecommunication fiber using externally written gratings," Electron. Lett. 26, 1270 (1990).
10. R. C. Youngquist, J. L. Brooks, and H. J. Shaw, "Two-mode fiber modal coupler," Opt. Lett. 9, 177 (1984).
11. F. Dios, A. B. Shvartsburg, D. Artigas, and F. Canal, "Nonlinear resonant conversion of modes in optical waveguides," Opt. Commun. 118, 28 (1995).
12. K. S. Lee and T. Erdogan, "Fiber mode coupling in transmissive and reflective tilted fiber gratings," Appl. Opt. 39, 1394 (2000).
13. Y. V. Kartashov, V. A. Vysloukh, and L. Torner, "Resonant mode oscillations in modulated waveguiding structures," Phys. Rev. Lett. 99, 233903 (2007).
14. K. Shandarova, C. E. Rüter, D. Kip, K. G. Makris, D. N. Christodoulides, O. Peleg, and M. Segev," Experimental observation of Rabi oscillations in photonic lattices," Phys. Rev. Lett. 102, 123905 (2009).
15. C. D. Poole, C. D. Townsend, and K. T. Nelson, "Helical-grating two-mode fiber spatial-mode coupler," J. Lightwave Technol. 9, 598 (1991).
16. K. S. Lee and T. Erdogan, "Fiber mode conversion with tilted gratings in an optical fiber," J. Opt. Soc. Am. A 18, 1176 (2001).
17. D. McGloin, N. B. Simpson, M. J. Padgett, "Transfer of orbital angular momentum from a stressed fiber-optic waveguide to a light beam," Appl. Opt. 37, 469 (1998).
18. C. N. Alexeyev and M. A. Yavorsky, "Generation and conversion of optical vortices in long-period helical core optical fibers," Phys. Rev. A 78, 043828 (2008).
19. C. N. Alexeyev, T. A. Fadeyeva, B. P. Lapin, and M. A. Yavorsky, "Generation of optical vortices in layered helical waveguides," Phys. Rev. A 83, 063820 (2011).

20. Y. V. Kartashov, V. A. Vysloukh, L. Torner, "Dynamics of topological light states in spiraling structures," Opt. Let. 38, 3414 (2013).
21. K. P. Marzlin and W. Zhang, "Quantized circular motion of a trapped Bose-Einstein condensate: coherent rotation and vortices," Phys. Rev. A 57, 4761 (1998).
22. V. I. Yukalov, K. P. Marzlin, and E. P. Yukalova, "Resonant generation of topological modes in trapped Bose-Einstein gases," Phys. Rev. A 69, 023620 (2004).
23. K. Staliunas and V. J. Sanchez-Morcillo, Transverse patterns in nonlinear optical resonators, Springer tracts in modern physics, book 183 (Springer, Berlin, 2003).
24. M. Brambilla, F. Battipede, L. A. Lugiato, V. Penna, F. Prati, C. Tamm, and C. O. Weiss, "Transverse laser patterns. 1. Phase singularity crystals," Phys. Rev. A 43, 5090 (1991).

# Figure captions

Figure 1. The weights of the $w_0, w_1$, and $w_2$ modes versus propagation distance in the linear medium at $\mu_b = 0.0333$, $\mu_d = 0.0167$ at resonant conditions. In (a) only $w_0$ mode was launched into the waveguide at $\xi = 0$, in (b) – only $w_1$ mode, and in (c) – only $w_2$ mode.

Figure 2. (a)-(c) Evolution dynamics in linear medium corresponding to Figs. 1(a)-1(c). (d)-(e) Evolution dynamics in nonlinear medium for different nonlinearity coefficients. In all cases $\mu_b = 0.0333$, $\mu_d = 0.0167$, $\delta\Omega_b = 0$, $\delta\Omega_d = 0$ and total propagation distance is $\xi = 460$.

Figure 3. (a) Maximal mode weights and (b) distances where weights of different modes acquire their maximal values versus relative detuning $\delta\Omega_b$ (in percent) of the bending frequency from its resonant value at $\mu_b = 0.0333$, $\mu_d = 0.0167$, $\delta\Omega_d = 0$. (c) Maximal mode weights versus $\mu_d$ at $\mu_b = 0.025$, $\delta\Omega_b = 0, \delta\Omega_d = 0$. In all cases $\sigma = 0$ and at $\xi = 0$ only $w_0$ mode was launched into waveguide.

Figure 4. Weights of the $w_1$ and $w_2$ modes at $\xi = 232$ versus relative detuning $\delta\Omega_b$ at $\sigma = -0.075$ (a) and $\sigma = -0.1$ (b) and $\mu_b = 0.0333$, $\mu_d = 0.0167$, $\delta\Omega_d = 0$. (c) Weights of the $w_1$ and $w_2$ modes at $\xi = 232$ versus nonlinearity coefficient at $\mu_b = 0.0333$, $\mu_d = 0.0167$, $\delta\Omega_b = 0$, and $\delta\Omega_d = 0$. Circles correspond to propagation dynamics shown in Fig. 2(a) and 2(d)-2(f). In all cases at $\xi = 0$ only $w_0$ mode was launched into waveguide.

Figure 5. (a) Linear modes of two-dimensional Gaussian waveguide. Black line shows potential well, while each mode profile is shifted in the vertical direction (by $-b_k^m$) to show the position of corresponding energy levels inside the well. Here $m$ denotes topological charge and $k$ is the number of radial nodes. (b) The weights of the $w_1^0$ and $w_0^1$ modes at $\xi = 181$ versus nonlinearity coefficient at $\mu_r = 0.002$, $\mu_d = 0.0081$, $\Omega_r = 3.357$, and $\Omega_d = 5.406$.

Figure 6. Transverse intensity distributions in linear medium showing (a) input fundamental mode at $\xi = 0$, (b) dynamically generated vortex mode at $\xi = 257$, $\mu_r = 0.002$, $\mu_d = 0$, and (c) dynamically generated mode with one radial node at $\xi = 257$, $\mu_r = 0$, $\mu_d = 0.0081$. Panels (d)-(e) show the impact of weak defocusing nonlinearity on the output intensity distribution at $\xi = 181$ in the case of complex two-frequency modulation with $\mu_r = 0.002$, $\mu_d = 0.0081$. Panels (d)-(e) correspond to circles in Fig. 5(b). In (d) the modes $w_1^0$ and $w_0^1$ have nearly equal weights, while with increase of nonlinearity the mode $w_1^0$ may be partially (e) or completely (f) suppressed. In all cases $\Omega_r = 3.357$, $\Omega_d = 5.406$.

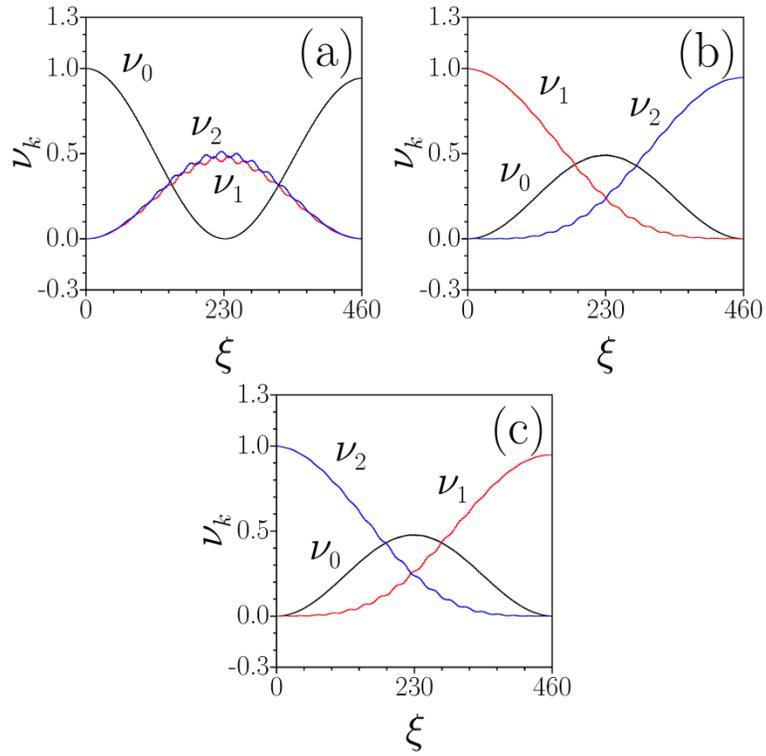

Figure 1. The weights of the $w_0, w_1$, and $w_2$ modes versus propagation distance in the linear medium at $\mu_b = 0.0333$, $\mu_d = 0.0167$ at resonant conditions. In (a) only $w_0$ mode was launched into the waveguide at $\xi = 0$, in (b) – only $w_1$ mode, and in (c) – only $w_2$ mode.

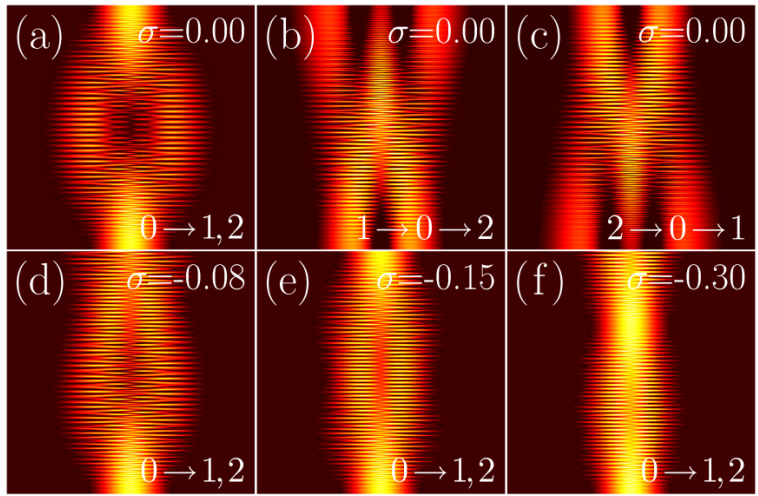

Figure 2. (a)-(c) Evolution dynamics in linear medium corresponding to Figs. 1(a)-1(c). (d)-(e) Evolution dynamics in nonlinear medium for different nonlinearity coefficients. In all cases $\mu_b = 0.0333$, $\mu_d = 0.0167$, $\delta\Omega_b = 0$, $\delta\Omega_d = 0$ and total propagation distance is $\xi = 460$.

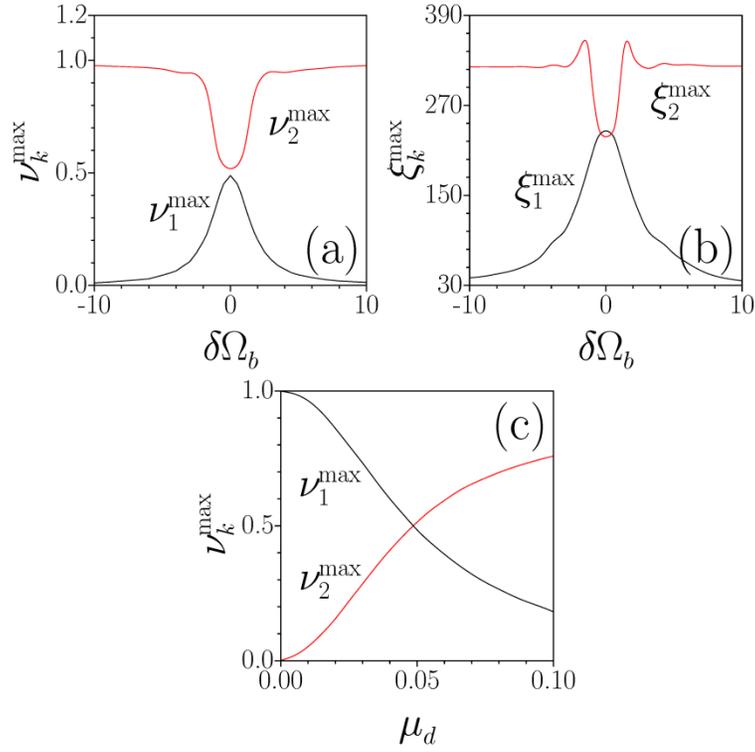

Figure 3. (a) Maximal mode weights and (b) distances where weights of different modes acquire their maximal values versus relative detuning $\delta\Omega_b$ (in percent) of the bending frequency from its resonant value at $\mu_b = 0.0333$, $\mu_d = 0.0167$, $\delta\Omega_d = 0$. (c) Maximal mode weights versus $\mu_d$ at $\mu_b = 0.025$, $\delta\Omega_b = 0$, $\delta\Omega_d = 0$. In all cases $\sigma = 0$ and at $\xi = 0$ only $w_0$ mode was launched into waveguide.

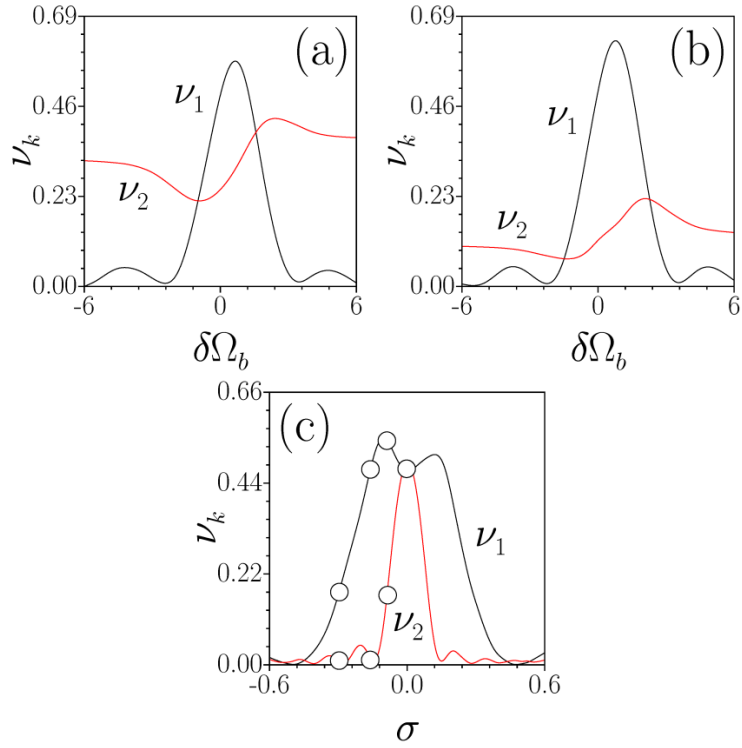

Figure 4. Weights of the $w_1$ and $w_2$ modes at $\xi=232$ versus relative detuning $\delta\Omega_b$ at $\sigma=-0.075$ (a) and $\sigma=-0.1$ (b) and $\mu_b=0.0333$, $\mu_d=0.0167$, $\delta\Omega_d=0$. (c) Weights of the $w_1$ and $w_2$ modes at $\xi=232$ versus nonlinearity coefficient at $\mu_b=0.0333$, $\mu_d=0.0167$, $\delta\Omega_b=0$, and $\delta\Omega_d=0$. Circles correspond to propagation dynamics shown in Fig. 2(a) and 2(d)-2(f). In all cases at $\xi=0$ only $w_0$ mode was launched into waveguide.

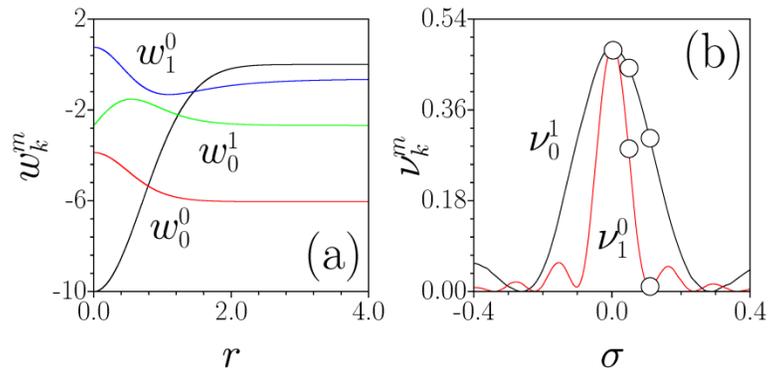

Figure 5. (a) Linear modes of two-dimensional Gaussian waveguide. Black line shows potential well, while each mode profile is shifted in the vertical direction (by $-b_k^m$) to show the position of corresponding energy levels inside the well. Here $m$ denotes topological charge and $k$ is the number of radial nodes. (b) The weights of the $w_1^0$ and $w_0^1$ modes at $\xi=181$ versus nonlinearity coefficient at $\mu_r=0.002$, $\mu_d=0.0081$, $\Omega_r=3.357$, and $\Omega_d=5.406$.

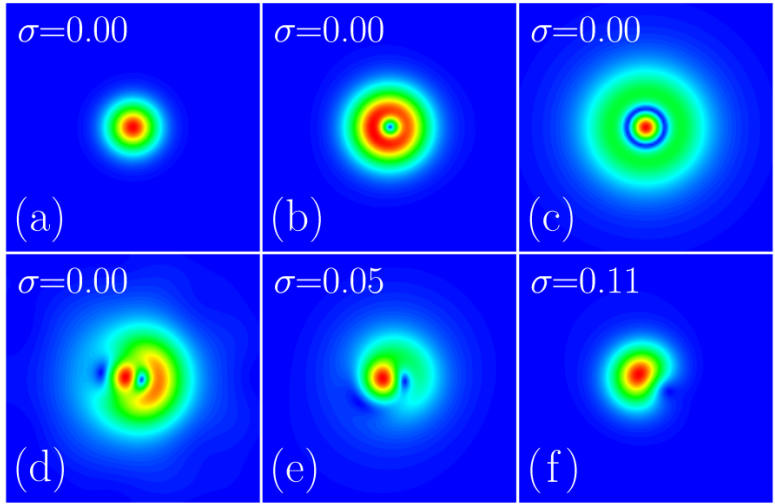

Figure 6. Transverse intensity distributions in linear medium showing (a) input fundamental mode at $\xi=0$, (b) dynamically generated vortex mode at $\xi=257$, $\mu_r=0.002$, $\mu_d=0$, and (c) dynamically generated mode with one radial node at $\xi=257$, $\mu_r=0$, $\mu_d=0.0081$. Panels (d)-(e) show the impact of weak defocusing nonlinearity on the output intensity distribution at $\xi=181$ in the case of complex two-frequency modulation with $\mu_r=0.002$, $\mu_d=0.0081$. Panels (d)-(e) correspond to circles in Fig. 5(b). In (d) the modes $w_1^0$ and $w_0^1$ have nearly equal weights, while with increase of nonlinearity the mode $w_1^0$ may be partially (e) or completely (f) suppressed. In all cases $\Omega_r=3.357$, $\Omega_d=5.406$.